\documentclass[superscriptaddress,twocolumn,secnumarabic,
 amssymb,amsmath,nobibnotes,aps,prd,showkeys,showpacs,nofootinbib]{revtex4}

\usepackage{graphicx}
\usepackage{epsf}
\usepackage{bm}%
\newcommand{\be}{\begin{equation}}
\newcommand{\ee}{\end{equation}}

\newcommand{\CC}{\Lambda}

\newcommand{\Omm}{\Omega_m}

\newcommand{\OL}{\Omega_{\Lambda}}

\newcommand{\rmo}{\rho_{m0}}
\newcommand{\rmm}{\rho_{m}}

\newcommand{\mincir}{\raise
-3.truept\hbox{\rlap{\hbox{$\sim$}}\raise4.truept\hbox{$<$}\ }}
\newcommand{\magcir}{\raise
-3.truept\hbox{\rlap{\hbox{$\sim$}}\raise4.truept\hbox{$>$}\ }}

\newcommand{\rL}{\rho_{\Lambda}}
\newcommand{\rM}{\rho_m}
\newcommand{\pM}{P_m}
\newcommand{\pL}{P_{\CC}}

\begin{document}
\hyphenation{tho-rou-ghly in-te-gra-ting e-vol-ving con-si-de-ring
ta-king me-tho-do-lo-gy fi-gu-re}

\title{The spherical collapse model in time varying vacuum cosmologies}

\author{Spyros Basilakos}
\affiliation{Academy of Athens, Research Center for Astronomy and Applied Mathematics,
 Soranou Efesiou 4, 11527, Athens, Greece}

\author{Manolis Plionis}
\affiliation{Institute of Astronomy \& Astrophysics, National Observatory of Athens,
Thessio 11810, Athens, Greece and
\\Instituto Nacional de Astrof\'isica, \'Optica y Electr\'onica, 72000 Puebla, Mexico}

\author{Joan Sol\`a}

\affiliation{High Energy Physics Group, Dept. Estructura i
Constituents de la Mat\`eria, Universitat de Barcelona, Diagonal
647, 08028 Barcelona, Catalonia, Spain and
\\Institut de Ci\`encies del Cosmos, UB, Barcelona}

\begin{abstract}
We investigate the virialization of cosmic structures in the
framework of flat FLRW cosmological models, in which the vacuum
energy density evolves with time. In particular, our analysis
focuses on the study of spherical matter perturbations, as they
decouple from the background expansion, ``turn
around'' and finally collapse. We generalize the spherical collapse
model in the case when the vacuum energy is a running function of
the Hubble rate, $\CC=\CC(H)$. A particularly well motivated model
of this type is the so-called quantum field vacuum, in which
$\CC(H)$ is a quadratic function, $\CC(H)=n_0+n_2\,H^2$, with
$n_0\neq 0$. This model was previously studied by our team
using the latest high quality cosmological data to constrain its
free parameters, as well as the predicted cluster formation rate. It
turns out that the corresponding Hubble expansion history resembles
that of the traditional $\Lambda$CDM cosmology. We use this
$\Lambda(t)$CDM framework to illustrate the fact that the properties
of the spherical collapse model (virial density, collapse factor,
etc.) depend on the choice of the considered vacuum energy
(homogeneous or clustered). In particular, if the distribution of
the vacuum energy is clustered, then, under specific conditions, we
can produce more concentrated structures with respect to the
homogeneous vacuum energy case.

\end{abstract}
\maketitle

\section{Introduction}
Several cosmological observations (supernovae type Ia, CMB, galaxy
clustering, etc.) have converged to a paradigm of a cosmic expansion history
that involves a spatially flat geometry and a recently initiated
accelerated expansion of the universe (see
\cite{Teg04,Spergel07,essence,Kowal08,Hic09,komatsu08} and
references therein). From a theoretical point of view, an easy way
to explain such expansion is to consider an additional energy
component, usually called dark energy (DE) with negative pressure,
that dominates the universe at late times. The simplest DE candidate
corresponds to the cosmological constant (see
\cite{Weinberg89,Peebles03,Pad03} for reviews).
An elegant model that fits accurately the current observational data,
is the spatially flat concordance $\Lambda$CDM model, which
includes cold dark matter (DM) and a cosmological constant, $\Lambda$.
However, the $\Lambda$ model suffers, among other
\cite{Peri08}, form two fundamental problems: (a) {\it The ``old''
cosmological constant problem} (or {\it fine tuning problem}) i.e.,
the fact that the observed value of the vacuum energy density
($\rho_{\Lambda}=\Lambda c^{2}/8\pi G\simeq 10^{-47}\,GeV^4$) is
many orders of magnitude below the value found using quantum field
theory \cite{Weinberg89}, and (b) {\it the coincidence
problem}\,\cite{coincidence} i.e., the fact that the matter energy
density and the vacuum energy density are of the same order (just
prior to the present epoch), despite the fact that the former is a
rapidly decreasing function of time while the latter is
stationary.

Many authors have attempted to solve the above problems (see
\cite{Peebles03,Pad03,Egan08} and references therein), the key
approach being the replacement of the constant vacuum energy either
with a time evolving DE (quintessence and the
like\,\cite{Peebles03}), or alternatively with a time varying vacuum
energy density, $\rho_{\Lambda}(t)$
\cite{Shap00,Shap02,Reuter00,Babic02,Grande06,Bas09b}. In the
original scalar field models\,\cite{Dolgov82} and later in the
quintessence context, one can ad-hoc introduce an adjusting or
tracker scalar field $\phi$\,\cite{Caldwell98} (different from the
usual SM Higgs field), rolling down the potential energy $V(\phi)$,
which could mimic the DE \cite{Peebles03,Pad03,Jassal,SR,Xin,SVJ}.
However, it was realized that the idea of a scalar field rolling
down some suitable potential does not really solve the problem
because $\phi$ has to be some high energy field of a Grand Unified
Theory (GUT), and this leads to an unnaturally small value of its
mass, namely one which is beyond all conceivable standards in
Particle Physics. As an example, utilizing the simplest form for the
potential of the scalar field, $V(\phi)=m_{\phi}^2\,\phi{^2}/2$, the
present value of the associated vacuum energy density is
$\rho_{\CC}=\langle V(\phi) \rangle\sim 10^{-11}\,eV^4$, so for
$\langle \phi\rangle$ of order of a typical GUT scale near the
Planck mass, $M_P\sim 10^{19}$ GeV, the corresponding mass of $\phi$
is expected in the ballpark of $m_{\phi}\sim\,H_0\sim 10^{-33}\,eV$.

Notice that the presence of such a tiny mass scale in scalar
field models of DE is generally expected also on the basis of
structure formation arguments\,\cite{Mota04,Nunes06}; namely from
the fact that the DE
perturbations 
seem to play an insignificant role in structure formation for scales
well below the sound horizon. The main reason for this homogeneity
of the DE is the flatness of the potential, which is necessary to
produce a cosmic acceleration. Being the mass associated to the
scalar field fluctuation, proportional to the second derivative of
the potential itself, it follows that $m_{\phi}$ will be very small
and one expects that the magnitude of DE fluctuations induced by
$\phi$ should be appreciable only on length scales of the order of
the horizon. Thus, equating the spatial scale of these fluctuations
to the Compton wavelength of $\phi$ (hence to the inverse of its
mass) it follows once more that $m_{\phi}\lesssim\,H_0\sim
10^{-33}\,eV$. All in all, it appears that the problem that one is
creating along with the introduction of $\phi$ is far more worrisome
than the problem one is intending to solve, for one is postulating a
mass scale which is $30$ orders of magnitude below the mass scale
associated to the value of the vacuum energy density
($m_{\Lambda}\equiv\rho_{\Lambda}^{1/4}\sim 2.3\times 10^{-3}\,eV$).

The analysis of the recent cosmological observations indicates that
the DE equation of state (EOS) parameter $w (\equiv P_{\rm DE}/\rho_{\rm DE})$
is close to $-1$ to within $\pm 10\%$, if it is assumed to be
constant \cite{Teg04,Spergel07,essence,Kowal08,Hic09,komatsu08},
whilst it is much more poorly constrained if it varies with
time\,\cite{essence}. More than two decades ago, Ozer \& Taha
\cite{Oze87} proposed a time varying $\Lambda$ as a possible
candidate to solve the two fundamental cosmological puzzles, see
also \cite{Free187,Carvalho92} and references therein. In this
cosmological paradigm, the dark energy EOS parameter $w$ is strictly
equal to -1, but the vacuum energy density (or $\Lambda$) does
evolve with time. Of course, the weak point in this approach 
is the
unknown functional form of $\Lambda(t)$, which is however also the
case for the vast majority of the DE models. Indeed, in the
aforementioned $\Lambda(t)$ models, the  evolution law is purely
phenomenological\,\cite{Overduin98}, without a concrete 
link to fundamental physics, such as the Quantum Field Theory
(QFT) in Curved Space time\,\cite{ParkerToms09}. As emphasized in
\cite{ShapSol09}, a completely consistent formulation along these
lines should eventually be developed, and such investigations could
well be at the heart of one the most important endeavors of
theoretical cosmology in the years to come. Therefore, the study of
cosmic perturbations in these models is very
important\,\,\cite{Grande08,GSFS10} as they might reveal surprises
not foreseen in the context of the scalar field models.
The new effects may have impact both on the cosmological
and the \textit{astrophysical} domains. While we recently analyzed
the potential implications on the former\,\cite{Bas09c}, here we
focus on the latter domain.

A pioneering QFT fundamental approach to variable $\CC$ models was
actually proposed long ago within the context of the renormalization
group (hereafter RG) in\, \cite{NelPan82,Toms83}. Later on, the
RG-running framework was further explored from different points of
view in \cite{Shap00,Shap02,Reuter00}, and a more systematic
presentation from the viewpoint of QFT in curved space-time by
employing the standard perturbation RG-techniques of Particle
Physics appeared in\,\cite{Shap02,Babic02}. Subsequent elaborations,
and comparison with the observational tests, confirmed the
phenomenological viability of this approach
\,\cite{RGTypeIa,SSS04,SS12,Fossil07,Bauer05,FSS09a,BFLWard}.

In the class of RG models we shall focus on, the vacuum energy
density is expected to vary with time according to the
law\,\cite{Shap02,RGTypeIa,SSS04,SS12,Fossil07}: $\CC=n_0+n_2\,H^2$
(hereafter called the $\Lambda_{RG}$ model or quantum field vacuum
model). As already mentioned, in Ref.\,\cite{Bas09c} we have
investigated thoroughly the global dynamics of this cosmological
model (together with various alternative $\Lambda(t)$ models), in
the light of the most recent cosmological data. However, a serious
issue here is how the main features of the largest collapsed cosmic
structures, i.e., galaxy clusters, are affected by a running vacuum
energy density. We have argued above that this problem can be
addressed in scalar field models of the DE, but only at the expense
of admitting extremely tiny mass scales which are uncommon in
Particle Physics. In this paper, we wish to further explore the
alternative option in which the DE component is a time evolving
cosmological term $\Lambda=\Lambda(t)$, and in this way to assess if
the clustering properties of the vacuum energy can shed some light
on the fundamental issue of structure formation.

The so called spherical collapse model\,\cite{Gunn72}, which has a
long history in cosmology, is a simple but still a fundamental tool
used to describe the growth of bound systems in the universe via
gravitation instability\,\cite{Pee93}. In the last decade many
authors have studied the small scale dynamics using this model
and found that the main features of the cosmic
structures (collapse factor, virial density, etc) can
potentially be affected by the presence of dark energy
\cite{Lahav91,Wang98,Iliev01,Lokas,Battye03,Maini03,Bas03,Wein03,
Mota04,manera,Horel05,Zeng05,Maor05,Perc05,Nunes06,Wang06,david,Fran08,
Basi07,Shaef08,Lee09,Abramo}. The aim of the present work is to
generalize the spherical collapse model within the variable $\Lambda_{RG}$ cosmological
model, in order to understand non-linear structure formation in
such cosmologies and investigate the differences with the respect to the 
expectations of the concordance $\Lambda$CDM cosmology.

The structure of the paper is as follows. The basic theoretical
elements of the problem are presented in section 2, where we
introduce [for a spatially flat Friedmann-Lema\^\i
tre-Robertson-Walker (FLRW) geometry] the basic cosmological
equations. In section 3  we generalize the virial theorem in the
case of the QFT $\Lambda(t)$ cosmological model. Section 4 outlines
the theoretical analysis of the spherical collapse model in which
the vacuum energy density varies with the cosmic time, and in
section 5 we compare the corresponding theoretical predictions of
the different models and present a first attempt to use
observational data to constrain the different models.
We draw our conclusions in section 6. In the appendix A we remind
the reader of some basic elements of the concordance $\Lambda$CDM
model in order to appreciate the fact that the $\Lambda_{RG}$
cosmology is an interesting extension of the standard model.
Finally, in appendix B we provide some basic mathematical
formulae, while in appendix C we provide accurate fitting
formulae for a few important parameters, ie., the density contrast at the
turn around redshift and at the epoch of virialization, which do not
have a simple fully analytical form.

\section{Cosmology with a time dependent vacuum}

The cosmological constant contribution to the curvature of
space-time is represented by the $\Lambda\,g_{\mu\nu}$ term on the
\textit{l.h.s.} of Einstein's equations.  The latter can be absorbed
on the \textit{r.h.s.} of these equations
\begin{equation}
R_{\mu \nu }-\frac{1}{2}g_{\mu \nu }R=8\pi G\ \tilde{T}_{\mu\nu}\,,
\label{EE}
\end{equation}
where the modified $\tilde{T}_{\mu\nu}$ is given by
$\tilde{T}_{\mu\nu}\equiv T_{\mu\nu}+g_{\mu\nu}\,\rL $. Here
$\rL=\CC/(8\pi G)$ is the vacuum energy density associated to the
presence of $\CC$ (with pressure $\pL=-\rL$), and $T_{\mu\nu}$ is
the ordinary energy-momentum tensor of isotropic matter and
radiation. Modeling the expanding universe as a perfect fluid with
velocity $4$-vector field $U_{\mu}$, we have
$T_{\mu\nu}=-\pM\,g_{\mu\nu}+(\rM+\pM)U_{\mu}U_{\nu}$, where $\rM$
is the proper isotropic pressure of matter-radiation and $\pM$ is
the corresponding pressure. Clearly the modified
$\tilde{T}_{\mu\nu}$ defined above takes the same form as
${T}_{\mu\nu}$ with $\tilde{\rho}=\rM+\rL$ and
$\tilde{p}=\pM+\pL=\pM-\rL$, that is
$\tilde{T}_{\mu\nu}=-\tilde{p}\,g_{\mu\nu}+(\tilde{\rho}+\tilde{p})U_{\mu}U_{\nu}$.
Explicitly,
\begin{equation}
\tilde{T}_{\mu\nu}= (\rL-\pM)\,g_{\mu\nu}+(\rM+\pM)U_{\mu}U_{\nu}\,.
\label{Tmunuideal}
\end{equation}
With this generalized energy-momentum tensor, and in the spatially
flat FLRW metric  $ds^{2}=dt^{2}-a^{2}(t)d\vec{x}^{2}$, the
gravitational field equations boil down to Friedmann's equation
\begin{equation}
H^{2}\equiv \left( \frac{\dot{a}}{a}\right) ^{2}=\frac{8\pi\,G }{3}%
\tilde{\rho}=\frac{8\pi\,G }{3}%
\left( \rM +\rL\right)\,,  \label{frie1}
\end{equation}
and the dynamical field equation for the scale factor:
\begin{equation}
\frac{\ddot{a}}{a}=-\frac{4\pi
G}{3}\,(\tilde{\rho}+3\,\tilde{p})=-\frac {4\pi
G}{3}\,(\rM+3\,\pM-2\,\rL)\,. \label{2FL}
\end{equation}
Let us next contemplate the possibility that $\rL=\rL(t)$ is a
function of the cosmic time. This is allowed by the Cosmological
Principle embodied in the FLRW metric. The Bianchi identities (which
ensure the covariance of the theory) then imply
$\bigtriangledown^{\mu}\,{\tilde{T}}_{\mu\nu}=0$. With the help of
the FLRW metric, the previous identity amounts to the following
generalized local conservation law:
\be \dot{\rho}_{m}+\dot{\rho_{\Lambda}}+ 3H(\rho_{m}+P_{m}+
\rho_{\Lambda}+P_{\Lambda})=0\,, \label{frie2} \ee
where the over-dot denotes derivative with respect to the cosmic
time.  The above equation can also be derived by combining eqs.
(\ref{frie1}) and (\ref{2FL}) since it is a first integral of the
equations of motion. Notice that we keep $G$ strictly constant, and
therefore the assumption  $\dot{\rho_{\Lambda}}\neq 0$ necessarily
requires some energy exchange between matter and vacuum, e.g.
through vacuum decay into matter, or vice versa\,\footnote{There
exists also the possibility that the vacuum is time evolving and
nevertheless non-interacting with matter. In this case, however,
either the DE has another component apart from $\CC$-- see the
$\CC$XCDM framework of\,\cite{Grande06} -- or Newton's coupling is
also time-varying, i.e. $\dot{G}\neq
0$\,\cite{SSS04,Fossil07,GSFS10,Letelier10}.}. This possibility was
first considered by M. Bronstein in a rather early
paper\,\cite{Bronstein33}.

Let us remark that the EOS of the vacuum energy density maintains
the usual form $P_{\Lambda}(t)=-\rho_{\Lambda}(t)=-\Lambda(t)/8\pi
G$ despite the fact that $\Lambda(t)$ evolves with time. In the
matter dominated epoch ($P_{m}=0$), eq.(\ref{frie2}) leads to the
following energy exchanging balance between matter and vacuum:
\begin{equation}
\dot{\rho}_{m}+3H\rho_{m}=-\dot{\rho_{\Lambda}}\,. \label{frie33}
\end{equation}
The second Friedmann's equation (\ref{2FL}) is formally unchanged by
the presence of a time-variable vacuum energy, and in the matter
epoch simply reads \be \frac{\ddot{a}}{a}=-\frac{4\pi
G}{3}\;(\rho_{m}-2\rho_{\Lambda}) \;. \label{ffrie2} \ee
At this point it is worth noticing that the effect of having a
variable cosmological term $\rL=\rL(t)$ cannot, in general, be
described by the simple parameterizations usually employed for the
effective EOS $w=w(t)$ of the DE, in which $w$ depends on two
parameters $(w_0,w_1)$, that can constrained using 
observations\,\cite{Jassal,quintEOS}. The effective EOS of a
variable vacuum model is in general more complicated. This is shown
in detail, with specific examples, in reference \cite{SS12}. In
particular, the vacuum models that we are going to consider cannot
be described with these simple parameterizations. Therefore, the
variable vacuum models must be studied on their own and constitute
an independent class of DE models. 

Combining equations (\ref{frie1}) and (\ref{frie33}), we find:
\begin{equation}
\dot{H}+\frac{3}{2} H^{2}=4\pi G\rho_{\Lambda}=\frac{\Lambda}{2}\,.
\label{frie34}
\end{equation}
If the vacuum term is negligible, $\Lambda \to 0$, then the solution
of eq. (\ref{frie34}) reduces to that of the Einstein-de Sitter
model, $H(t)=2/3t$, as it should. Similarly, the traditional
$\Lambda=$const. cosmology (or $\CC$CDM concordance model) also
follows directly by integrating eq.(\ref{frie34}) [see appendix A].
Finally, this same equation is also valid for $\Lambda=\Lambda(t)$
when matter and vacuum become coupled, and in this case a
supplementary equation for the time evolution of $\Lambda$ is needed
in order to unveil the dynamics of this model.  It is interesting to
mention here that the link in eq.(\ref{frie33}) between
$\dot{\rho}_m$ and $\dot{\Lambda}$ is important because interactions
between DM and DE could provide possible solutions to the
cosmological coincidence problem. This is the reason for which
several papers have been published recently in this area
\cite{Zim01,Grande06,Bas09b}, proposing either that the DE has
various interacting components or that the DE and DM could be
coupled. In the following subsections, we briefly introduce (for
more details see \cite{Bas09c}) the cosmological models used in this
study.

\subsection{The $\Lambda(t)$ model from quantum field theory}
In this scenario, we use the vacuum solution proposed
in\,\cite{Shap02,RGTypeIa,SSS04,SS12,Fossil07} using the
renormalization group (RG) in quantum field theory (hereafter
$\Lambda_{RG}$ model). The model is characterized by the evolution
law: $\CC(H)=n_0+n_2\,H^2$, in which both coefficients $n_0$ and $n_2$
are non-vanishing. An equivalent and more convenient parametrization
is
\be \label{RGlaw2} \Lambda(H)=\Lambda_0+ 3\gamma\,(H^{2}-H_0^2)\,,
\ee
where $\gamma$ is a constant, which can be positive or negative but
small: $|\gamma| \le 1/12\pi$ \cite{Bas09c}. It determines the
amount of running of $\Lambda(t)$. Note that the form (\ref{RGlaw2})
is convenient because the vacuum energy density is normalized to the
present value:
$\Lambda_0\equiv\Lambda(H_0)=3\Omega_{\Lambda}H^{2}_{0}$. As we will
see, all the equations and conditions derived below are equivalent
to those of the concordance $\Lambda$CDM cosmology (cf. appendix A) and
reduce exactly to them for $\gamma=0$.

From equations (\ref{frie34}) and (\ref{RGlaw2}) we can easily
derive the corresponding Hubble flow as a function of time \cite{Bas09c}:
\begin{equation}
\label{frie455} H(t)=H_{0}\,\sqrt{\frac{\OL-\gamma}{1-\gamma}} \;
\coth\left[\frac32\,H_{0}\sqrt{(\OL-\gamma)(1-\gamma)}\;t\right]\,.
\end{equation}
The scale factor of the universe, $a(t)$, evolves as
\begin{equation}\label{frie456}
a(t)=a_1\, \sinh^{\frac{2}{3(1-\gamma)}}
\left[\frac32\,H_{0}\sqrt{(\OL-\gamma)(1-\gamma)}\;t\right]\,,
\end{equation}
where
\begin{eqnarray}
\label{all2}
a_{1}=\left(\frac{\Omega_{m}}
{\Omega_{\Lambda}-\gamma}\right)^{\frac{1}{3(1-\gamma)}}\,.
\end{eqnarray}
Inverting eq.(\ref{frie456}) we determine the cosmic time as a
function of the scale factor:
\begin{equation}
t(a)=\frac{2}{3\,{\tilde{\Omega}_{\Lambda}}^{1/2}\,(1-\gamma)\,H_{0}
} {\rm sinh^{-1}} \left(\sqrt{ \frac{\tilde{\Omega}_{\Lambda}}
{\tilde{\Omega}_{m}}} \;a^{3(1-\gamma)/2} \right) \label{frie456t}
\end{equation}
where we have introduced
\begin{equation}
\label{otran1}
\tilde{\Omega}_{m}=\frac{\Omega_{m}}{1-\gamma}\,,\ \ \
\tilde{\Omega}_{\Lambda}=\frac{\Omega_{\Lambda}-\gamma}{1-\gamma}\,.
\end{equation}

Let us define the Hubble rate normalized to its current value, $E(a)=H(a)/H_0$.
Eliminating the cosmic time from equations (\ref{frie455}) and
(\ref{frie456}), one can prove that
\begin{eqnarray}
\label{norm11} E^2(a)&=&
\frac{\OL-\gamma}{1-\gamma}+\frac{\Omm}{1-\gamma}a^{-3\,(1-\gamma)}\nonumber\\
&=& 1+\Omm\, \frac{a^{-3\,(1-\gamma)}-1}{1-\gamma}\,,
\end{eqnarray}
or
\begin{equation}
\label{anorm11}
E^{2}(a)=\tilde{\Omega}_{\Lambda}+\tilde{\Omega}_{m}a^{-3(1-\gamma)}\,.
\end{equation}

Notice that both sets of cosmological parameters satisfy the
standard cosmic sum rule:
\begin{equation}
\label{otran1b}
\tilde{\Omega}_{m}+\tilde{\Omega}_{\Lambda}=1=\Omega_{m}+\Omega_{\Lambda}\,.
\end{equation}

Let us also consider the evolution of the matter and vacuum
energy densities in this model.
Starting from the conservation law [see eq.\ref{frie33}]
and utilizing eqs.(\ref{frie1}) and
(\ref{RGlaw2}),
we arrive at a simple differential equation for the
matter density,
\begin{equation}\label{rhomRG}
\dot{\rho}_m+3H\rmm=3\gamma H\rmm\,.
\end{equation}
Using $\dot{\rho}_m=aH d\rho_m/da$, we can trivially
integrate the previous equation in the scale factor variable,
yielding
\begin{equation}\label{mRG}
\rho_m(a) =\rmo\,a^{-3(1-\gamma)}\,,
\end{equation}
where $\rmo$ denotes the matter density at the present time ($a=1$),
and therefore $\Omega_m=\rmo/\rho_{c0}$, where
$\rho_{c0}=3H_0^2/(8\pi G)$ is the current critical density.
The previous equation can also be rewritten by
considering the instantaneous critical density $\rho_c(a)$  when the
scale factor is $a=a(t)$, i.e. $\rho_c(a)\equiv 3H^2(a)/(8\pi G)$.
In fact, defining $\Omega_m(a)\equiv{\rho_m(a)}/{\rho_c(a)}$ it is
easy to see, with the help of (\ref{mRG}) and the definition of
$E(a)$, that
\begin{equation}\label{effeom}
\Omega_m(a)=\frac{\Omega_{m}a^{-3(1-\gamma)}}{E^2(a)}\,.
\end{equation}
Finally, upon inserting (\ref{mRG}) in (\ref{frie33}) and integrating
once more in the scale factor variable, we arrive at the explicit
expression for the evolution of the vacuum energy density:
\begin{equation}
\label{CRG}
\CC(a)=\CC_0+8\pi G \;\frac{\gamma\,\rmo}{1-\gamma}\,\left[a^{-3(1-\gamma)}-1\right]\,.
\end{equation}
It is important to emphasize from eq.\,(\ref{mRG}) that the matter
density does no longer evolve as $\rho_m(a)=\rmo a^{-3}$, as it
presents a correction in the exponent. This is due to the fact that
matter is exchanging energy with the vacuum and this is reflected in
the corresponding behavior of $\Lambda(a)$ in eq.(\ref{CRG}). As
expected, for $\gamma\to 0$ ($\tilde{\Omega}_{m} \sim \Omega_{m},\;
\tilde{\Omega}_{\Lambda}\sim \Omega_{\Lambda}$) all the above
equations reduce to the canonical form within the concordance
model (for more details, see the appendix A). Clearly, the usual
$\Lambda$-cosmology is a particular solution of the $\Lambda_{RG}$
model with $\gamma$ strictly equal to 0. Throughout the rest of the
paper we shall employ the statistical results for this model
obtained in\,\cite{Bas09c}  from a simultaneous fit to the latest
SNIa+BAO+CMB data, namely
$\Omega_{m}=0.28^{+0.02}_{-0.01}$ (or $\tilde{\Omega}_{m}\simeq
0.281$) and $\gamma=0.002\pm 0.001$.

\section{Generalization of the virial theorem}
Recall that for systems with potential energy of the
form $U\propto R^n$ the
contribution to the virial condition is $2T-nU=0$, where $T$ is the
kinetic energy. For gravity, $n=-1$, whereas for constant vacuum
energy $n=2$. Indeed, from Newton's limit of Einstein's equations in
the presence of a $\Lambda$ term, i.e. Poisson equation
$\nabla^{2} \Phi=4\pi G (\rho_{m} -2\rho_{\Lambda})$, it follows
that the potential associated to a constant $\Lambda$ is
$\Phi_{\Lambda}=-(1/6)\,\Lambda\,R^2$. Thus, the nominal virial
condition when there is a constant vacuum energy reads
$2T+U_{G}-2U_{\Lambda}=0$, where $U_{G}$ and $U_{\Lambda}$ are the
potential energy and the vacuum potential energy, respectively, for
an isolated system. However, for non-constant $\Lambda$ this recipe
may not hold anymore.
Therefore, the link between the
time-dependent vacuum and matter is expected to modify non-trivially
the previous form of the virial theorem.

In this section, we generalize the virial theorem by taking into
account the presence of the coupling between the vacuum and the
matter energy densities, which leads to a variable $\Lambda(t)$. In
particular, we have to modify the well know Layzer-Irvine equation,
which describes the flow to virialization \cite{Pee93}. As we have
already stated in section 2.1 (see eq.\ref{frie33}), the
matter is exchanging energy with
the vacuum and this is reflected in the corresponding matter
continuity equation (\ref{rhomRG}). Of course, the continuity
equation holds for the total cosmic fluid. Since, we are interested
for cosmic structures which live in high density environments, it is
fair to consider that the corresponding inhomogeneous density
$n_{m}$ is far from the background homogeneous density,
$\rho_m$.

The total velocity of the fluid elements, ${\bf v}$, is given by:
${\bf \nabla}\cdot (n_{m} {\bf v})= 3H n_{m}+{\bf \nabla}\cdot
(n_{m} {\bf u})$, where ${\bf u}$ is the peculiar velocity. Note that
${\bf v}$ is the sum of the global Hubble expansion plus the line of sight
peculiar velocity, ${\bf u}$.
The continuity equation for $n_{m}$, in the
presence of a running vacuum energy ($\gamma\neq 0$), can be now
 modified as follows:
\be \label{lazer1} {\dot n_{m}}+3H n_{m}+{\bf \nabla}\cdot (n_{m}
{\bf u})=3\gamma H n_{m}\,. \ee
Notice that if the background would not be expanding ($H=0$), then
eq. (\ref{lazer1}) boils down to the standard non-relativistic
continuity equation for a fluid of particles flowing with a velocity
distribution ${\bf u}$. Based on a Newtonian formulation, the
acceleration due to gravity in comoving coordinates is \be
\label{lazer2} \frac{d(a {\bf u})}{dt}=-a{\bf \nabla} \Phi \;, \ee
where $\Phi$ is the gravitational potential. Multiplying
eq.(\ref{lazer2}) {\bf by} the quantity $a n_{m} {\bf u}$ and using the
continuity eq.(\ref{lazer1}) together with the overall
Poisson equation \cite{Pee93} $\nabla^{2} \Phi=4\pi G (n_{m}
-2\rho_{\Lambda})$, we can integrate over the volume in order to
define the generalized Layzer-Irvine equation. After some
calculations we arrive at \be \label{lazer3} \frac{d(a^{2}
T)}{dt}-3\gamma a^{2}H T= -a^{2}(\frac{d{\cal U}}{dt}+H {\cal
U})+6\gamma a^{2}H {\cal U} \ee where ${\cal U}=U_{G}-2U_{\Lambda}$,
\be T=\frac{1}{2}\int u^{2} n_{m} dV \;, \ee \be U_{G}=
-\frac{1}{2}G \int \int
\frac{n_{m}(x)n_{m}(x^{'})}{|x-x^{'}|}dVdV^{'} \ee and \be
U_{\Lambda}= -\frac{1}{2}G \int \int \frac{n_{m}(x)
\rho_{\Lambda}(x^{'})}{|x-x^{'}|}dVdV^{'}\;. \ee For a spherical
mass fluctuation $M=4\pi n_{m} R^{3}/3$, one can show that the above
potential energies become \be \label{Matener} U_{G}=-\frac{16\pi^{2}
G}{3} \int_{0}^{R} x^{4} \; n^{2}_{m}(x) dx=-\frac{3GM^{2}}{5R} \ee
and \be \label{Lener} U_{\Lambda}=-\frac{16\pi^{2} G}{3}
\int_{0}^{R} x^{4} \; \rho_{\Lambda}(x) n_{m}(x) dx=-\frac{\Lambda
M}{10}R^{2} \ee where the last equality holds for a homogeneous
vacuum energy $\Lambda=\Lambda(a)$ [see \cite{Horel05,Maor05} and
section 5].

Now, for a system that reaches the equilibrium (virial regime, and
$\dot{\cal U}=\dot{T}=0$) we derive the condition \be \label{lazer4}
(2-3\gamma)T+(1-6\gamma)(U_{G}-2U_{\Lambda})=0 \ee or \be
\frac{U_{G}-2U_{\Lambda}}{T}=-2\;\frac{1-3\gamma/2}{1-6\gamma}
=-2-9\gamma+{\cal O}(\gamma^{2}) \ee where the last equality is
valid for small values of $\gamma$. For $\gamma$ strictly
equal to zero we obviously recover the nominal virial theorem in
the concordance cosmology ($\frac{U_{G}-2U_{\Lambda}}{T}=-2$)
at it should.
In the case of
$\gamma \simeq 1/12\pi$\,\cite{Bas09c}, which corresponds to
receiving quantum effects on the $\Lambda$ running from
fields just at the Planck scale\,\cite{Shap02,RGTypeIa}, the above
ratio in the $\Lambda(t)$ cosmology deviates by $\sim 12\%$ with
respect to that of the $\Lambda$ cosmology. In practice, however,
$\gamma \simeq 1/12\pi$ is excluded by the latest fit to the
combined data, which yields a quite smaller value: $\gamma \simeq
0.002$,\cite{Bas09c}. This value corresponds to quantum
effects from GUT fields just one order of magnitude below $M_P$. In
this case, the deviation from the nominal virial condition is only
$\sim 1\%$.

\begin{figure}[ht]
         \centerline{\includegraphics[width=23pc] {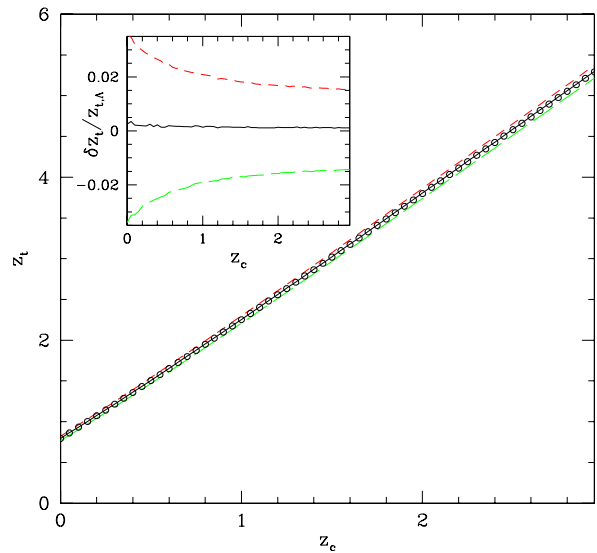}}
  \caption{The turn around redshift as a function
of the virial redshift, see eq.(\ref{ztzc}). The solid, long dashed
and short dashed lines
represent the $\Lambda_{RG}$ model, for $\gamma=0.002$, $1/12\pi$ and
$-1/12\pi$, respectively. The solid points correspond to the
concordance $\Lambda$CDM cosmology. {\em Inset Panel:} The relative
fractional difference between the three $\Lambda_{RG}$ models and the
concordance model.}
        \label{Figcondz}
 \end{figure}

\section{The spherical collapse model}
Despite its simplicity, the spherical collapse model\,\cite{Gunn72}
is still a powerful tool for understanding how a small spherical
patch of homogeneous overdensity forms a bound system via
gravitation instability --
for a review, see e.g.\,\cite{PadmanabhanBook}.
Technically speaking, the basic dynamical cosmological equation
\,(\ref{ffrie2}) is valid both for the entire universe and also for
a homogeneous spherical perturbation. In the last case, we just
replace the scale factor $a(t)$ with the radius $R(t)$, and we
obtain the so called Raychaudhuri equation:
\begin{equation}
\frac{\ddot{R}}{R}=-\frac{4\pi G}{3}[\rho_{ms}-2\rho_{\Lambda
s}]\,,\label{rayc}
\end{equation}
where $\rho_{ms}$ and $\rho_{\Lambda s}$ refer to the
corresponding values of the matter and vacuum energy densities in
the spherical patch susceptible of ulterior collapse.

In order to address the issue of how the time varying vacuum energy
itself affects the gravitationally bound systems (clusters of
galaxies), one has to deal in general with the following
three distinct scenarios, which have been considered
within different approaches in the literature\,
\cite{Lahav91,Wang98,Iliev01,Lokas,Battye03,Maini03,Bas03,Wein03,
manera,Mota04,Horel05,Zeng05,Maor05,Perc05,Nunes06,Wang06,david,Fran08,
Basi07,Shaef08,Lee09,Abramo}: (i) the situation in which the vacuum
energy remains homogeneous and only the corresponding matter
virializes; (ii) the case with clustered vacuum energy, but now
assuming that only the matter virializes; and, finally, (iii) the
case with clustered vacuum
energy, considering that the whole system virializes (both
matter and vacuum components). In this paper, we are going to focus
on scenarios (i) and (iii) within the framework of time varying
vacuum energy density.

From now on, we will call $a_{t}$ the value of the scale factor of
the universe where the spherical overdensity reaches its maximum
expansion (i.e. when $\dot R=0$) and $a_{c}$ the scale factor when
the sphere virializes, implying that a cosmic structure has formed.
Similarly, $R_{t}$ and $R_{c}$
stand for the corresponding radii of the
spherical overdensity, the former being the turnaround (or
``top hat'') value at the point of maximum size, and the latter
refers to the eventual situation when the sphere has already
collapsed and virialized. Note that due to the coupling between the
time-dependent vacuum and the matter component one would expect that
the matter density in the spherical region should obey the same
power law as the background matter $\rho_{m}(a)\propto
a^{-3(1-\gamma)}$ (see eq.\ref{mRG}). Thus, $\rho_{ms}\propto
R^{-3(1-\gamma)}$ denotes the matter density in the spherical patch.
Analogously, the vacuum energy density in the spherical
region, $\rho_{\Lambda s}$ will take the form (\ref{CRG}) with
appropriate replacement of the scale factor with $R$, see further below.

From the theoretical point of view,
the time needed for a spherical shell to re-collapse
is twice the turn-around time, $t_{f}\simeq 2t_{t}$, which implies
that (see eq.\ref{frie456t}):
\begin{equation}
{\rm sinh^{-1}}\left[\sqrt{ \tilde{r}_{0} \;a_{c}^{3(1-\gamma)}}
\right]\simeq 2{\rm sinh^{-1}}\left[\sqrt{\tilde{r}_{0} \;
a_{t}^{3(1-\gamma)}} \right] \;,\label{tim}
\end{equation}
where 
$\tilde{r}_0=\tilde{\Omega}_{\Lambda}/\tilde{\Omega}_{m}$.

In the main panel of Figure 1, we present the turnaround redshift,
$z_t=(1-a_t)/a_t$, as a function of the virial redshift
$z_c=(1-a_c)/a_c$ for our $\Lambda_{RG}$ model with
$\gamma=0.002$ (continuous line),
$\gamma=1/12\pi$ and $\gamma=-1/12\pi$ (long and short dashed lines,
respectively). The concordance $\Lambda$ cosmology is indicated by
empty points. The relative fractional differences between the $\Lambda_{RG}$ models
and the concordance $\Lambda$ model are
extremely small, a fact which can be appreciated in the inset panel of Fig.1.

As it is evident there is a tight correlation between the
two redshifts, which for our $\Lambda_{RG}$ model ($\gamma=0.002$) it is
given by:
\begin{equation}\label{ztzc}
z_{t}\simeq 1.532z_{c}+0.751\;\;.
\end{equation}

As an example, assuming that galaxy clusters have virialized at the
present time, $z_{c}\simeq 0$, the turn around epoch takes place at
$z_{t}\simeq 0.75$ (or $a_{t}\sim 0.57$). On the other hand,
considering that clusters have formed prior to the epoch of
$z_{c}\sim 1.6$ ($a_{c}\sim 0.38$), in which the most distant
cluster has been found \cite{Pap10}, the turn around epoch is not
really affected by the vacuum energy component, i.e. $z_{t}\sim 3.2$
(or $a_{t} \sim 0.24$). This is to be expected, due to the fact that
at large redshifts matter dominates the Hubble expansion. It is
worth noting that, at high redshifts, the ratio between the scale factors
approaches  the Einstein-de Sitter
($\tilde{\Omega}_{m}=\Omega_{m}=1$) value:
${a_{c}}/{a_{t}}=(1+z_{t})/(1+z_{c})=2^{2/3}$.

Performing the convenient transformations into dimensionless
variables
\begin{equation}
\label{trann}
x=\frac{a}{a_{t}}\;\;\;\; {\rm and}\;\;\;\;
y=\frac{R}{R_{t}} \;\;\;,
\end{equation}
the evolution of the scale factor of the background and
of the perturbation (see eqs.\ref{frie1}, \ref{anorm11}
and \ref{rayc}) are
governed respectively by the following two equations:
\begin{equation}\label{xdot}
\left(\frac{{\dot x}}{x}\right)^{2}=
H_{t}^{2}\Omega_{m,t} \left[x^{-3(1-\gamma)}+
\frac{\rho_{\Lambda}}{\rho_{m,t}}\right]
\end{equation}
and
\begin{equation}\label{yddot}
\frac{{\ddot y}}{y}=-\frac{H_{t}^{2}\Omega_{ m,t}}{2} \left[
\frac{\zeta}{y^{3(1-\gamma)}}-2\frac{\rho_{\Lambda s}}{\rho_{m,
t}}\right]\,,
\end{equation}
where $H^{2}_{t}\Omega_{ m,t}=\frac{8 \pi G}{3}\rho_{m, t}$,
$\Omega_{m,t}$ is the matter density parameter at the turn around
epoch (see eq.\ref{effeom})\footnote{We set $\Omega_{m, t}\equiv
\Omega_{m}(a_{t})$, with $\Omega_{m}(a)$ given by
eq.(\ref{effeom}).}. Note that we have parametrized the matter
density in the spherical region at the turn around time $\rho_{ms,
t}=\zeta \rho_{m, t}$, with respect to the background matter density
at the same epoch $\rho_{m, t}$. The parameter $\zeta$
is referred to as the density contrast at the turnaround point. In
replacing the quantity $\rho_{ms}$ in eq.(\ref{rayc}) we have
utilized the following relation:
\begin{equation}
\rho_{ms}=\rho_{ms, t} \left(\frac{R}{R_{t}}\right)^{-3(1-\gamma)}=
\frac{\zeta \rho_{m, t}}{y^{3(1-\gamma)}} \;.
\end{equation}
Therefore, using
eqs.(\ref{otran1},\ref{CRG}) we find
that \be
\label{Ixx}
\frac{\rho_{\Lambda}}{\rho_{\Lambda, t}}=I(x)=
\frac{1+\gamma \tilde{r}_{0} a^{-3(1-\gamma)}_{t} x^{-3(1-\gamma)} }
{1+\gamma \tilde{r}_{0} a^{-3(1-\gamma)}_{t} } \;,
\ee
$\rho_{\Lambda, t}$ being the vacuum energy density at the turn
around epoch and $\Omega_{\Lambda,t}=1- \Omega_{m,t}$ is the
effective vacuum density parameter at the same time (for definition
see eq.\ref{effeom}). Inserting now eq.(\ref{Ixx}) into
eq.(\ref{xdot}), we finally obtain
\begin{equation}
\label{xdot1}
{\dot x}^{2}=
H_{t}^{2}\Omega_{m,t} \left[x^{-1+3\gamma}+rx^{2}I(x)\right]
\end{equation}
where
\begin{equation}
r=\frac{\rho_{\Lambda, t}}{\rho_{m, t}}
=\frac{\Omega_{\Lambda}}{\Omega_{m}}a^{3(1-\gamma)}_{t}+
\frac{\gamma}{1-\gamma}\left[1-a^{3(1-\gamma)}_{t}\right]
\end{equation}
(to derive the latter equality we have used eqs.17,19,21).

Of course for the Einstein-de Sitter case
($\tilde{\Omega}_{m}=\Omega_{m}=1$ and $\gamma=0$) the solution of
the system formed by eq.(\ref{xdot}) and eq.(\ref{yddot}) reduces to
the well known value of the density contrast at the turnaround
point: $\zeta=\left(\frac{3 \pi}{4}\right)^{2}$, as it should.
Within this framework, utilizing both the virial theorem (see
eq.\ref{lazer4}) and the energy conservation
($T_{c}+U_{G,c}+U_{\Lambda,c}= U_{G,t}+U_{\Lambda,t}$) at the
collapse time and at the turn around epoch respectively we derive
the following useful relation: \be \label{energies}
q_{1}U_{G,c}+q_{2}U_{\Lambda,c}=U_{G,t}+U_{\Lambda,t} \ee where \be
\label{q12} q_{1}(\gamma)=\frac{1+3\gamma}{2-3\gamma} \;\;\;\;
q_{2}(\gamma)=\frac{4-15\gamma}{2-3\gamma} \ . \ee

 \begin{figure}[ht]
         \centerline{\includegraphics[width=23pc] {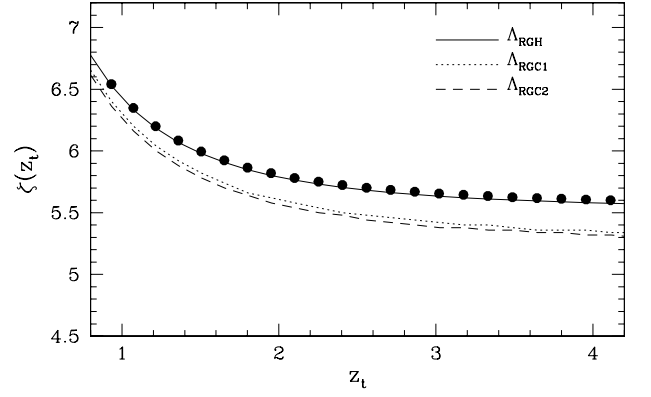}}
  \caption{The density contrast at the turn around epoch,
$\zeta$ as a function of the turn around redshift.
The lines represent the following cosmological models: (a)
$\Lambda_{RGH}$ (solid), (b)
$\Lambda_{RGC1}$ (dot line, $\gamma_{s}=0.002$) and (c)
$\Lambda_{RGC2}$ (dashed line, $\gamma_{s}=-0.002$).
The points represent the concordance $\Lambda$CDM cosmology.
}
        \label{Figcondz2}
 \end{figure}

\subsection{Homogeneous Vacuum Energy}
In this section, we consider that the vacuum energy component of the
scale of galaxy clusters can be treated as being homogeneous:
$\rho_{\Lambda s}(a)=\rho_{\Lambda}(a)=\Lambda(a)/8\pi G$ (hereafter
the $\Lambda_{RGH}$ model). Therefore, inserting eq.(\ref{Ixx}) into
eq.(\ref{yddot}), we obtain
\begin{equation}
\label{yddot1} {\ddot y}=-\frac{H_{t}^{2}\Omega_{ m,t}}{2}
\left[ \frac{\zeta}{y^{2-3\gamma}}-2r y I(x)\right]
\end{equation}

The numerical solution for the $\zeta$ parameter is provided by
integrating the main system of differential equations,
(eqs.\ref{xdot1} and \ref{yddot1}), using the boundary
conditions: $({\rm  d} y/{\rm d} x)=0$ and $y=1$ at $x=1$.
Following the methodology of \cite{Wang98} and \cite{Lee09} we
provide in the appendix C a reasonably accurate
fitting formula for $\zeta$, as a function
of the main cosmological parameters.

Using now, the combined equation (\ref{energies}) for the potential
energies (see eqs.\ref{Matener},\ref{Lener})\footnote{In view of the
fact that $U_{\Lambda}=-\Lambda(a)MR^{2}/10$, the time dependence of
the vacuum energy density seems to create a problem since the total
energy of the bound system is not conserved. However, one can show
that if the value of $|\gamma|$ is less that 0.01 then the problem
of energy conservation does not really affect the virialization
process and thus eq.(\ref{energies}) remains a good approximation.},
we obtain a cubic equation that relates the ratio between the virial
($R_{c}$) to the turn-around outer radius ($R_{t}$), the so called
collapse factor ($\lambda=R_{c}/R_{t}$):
\begin{equation}
\label{homvirial} q_{2}(\gamma)n_{c}
\lambda^{3}-(2+n_{t})\lambda+2q_{1}(\gamma)=0\,,
\end{equation}
where
\begin{equation}
n_{c}=\frac{\Lambda(a_{c})}{4\pi G \rho_{m, t} \zeta}=
n_{0}+
\frac{2\gamma a^{3(1-\gamma)}_{t}}{\zeta (1-\gamma)}\;
\left[a^{-3(1-\gamma)}_{c}-1\right]
\end{equation}
and
\begin{equation}
n_{t}=\frac{\Lambda(a_{t})}{4\pi G \rho_{m, t} \zeta}=
n_{0}+\frac{2\gamma a^{3(1-\gamma)}_{t}}{\zeta (1-\gamma)}
\;\left[a^{-3(1-\gamma)}_{t}-1\right]
\end{equation}
with
\be \label{noo}
n_{0}=\frac{2 \Omega_{\Lambda}
a^{3(1-\gamma)}_{t} } {\Omega_{m} \zeta}
 \;\;.
\ee
Finally, solving the cubic eq.(\ref{homvirial}), we calculate the
collapse factor (see appendix B). In the case of $\gamma=0$
($n_{c}=n_{t}=n_{0}$), the above expressions take the usual form of
the $\Lambda$ cosmology (see \cite{Bas03}, \cite{Lahav91}), as expected.
Obviously, for the Einstein-de Sitter model ($\Omega_{m}=1$,
$\gamma=0$) we have $\Omega_{\Lambda}=0$ and all coefficients vanish
$n_{c}=n_{t}=n_{0}=0$, so that eq.(\ref{homvirial}) boils down to
$\lambda=q_1(0)=1/2$, see eq.\,(\ref{q12}).

In this framework, the density contrast at the virialization epoch
is given by:
\begin{equation}\label{deltavir}
\Delta_{vir}=\frac{\rho_{ms, c}}{\rho_{m, c}}=
\frac{\zeta}{\lambda^{3}} \left(\frac{a_{c}}{a_{t}}\right)^{3}\;\;,
\end{equation}
where $\rho_{ms, c}$ is the matter density in the virialized structure
and $\rho_{m, c}$ is the background matter density
at the same epoch.
Following the notations of \cite{Wein03,Kita96,shaw}, we again
provide in appendix C a fitting formula for $\Delta_{vir}$
(within a physical range of cosmological parameters:
$0\le \gamma \le 0.01$).
Notice that the Einstein-de
Sitter value for $\Delta_{vir}$ is precisely $18\pi^{2}$, and
was
factorized in (\ref{aproxf})\,\footnote{This value
ensues from the Einstein de-Sitter value of $\zeta=(3\pi/4)^2$ multiplied by
$2^5$. Indeed, the sphere contracts a factor of $2$ from the
turnaround point to virialization, and the background scale factor
increases $a_{c}/a_{t}=2^{2/3}$, thus $\left(2\times
2^{2/3}\right)^3=2^5$.}.

 \begin{figure}[ht]
         \centerline{\includegraphics[width=23pc] {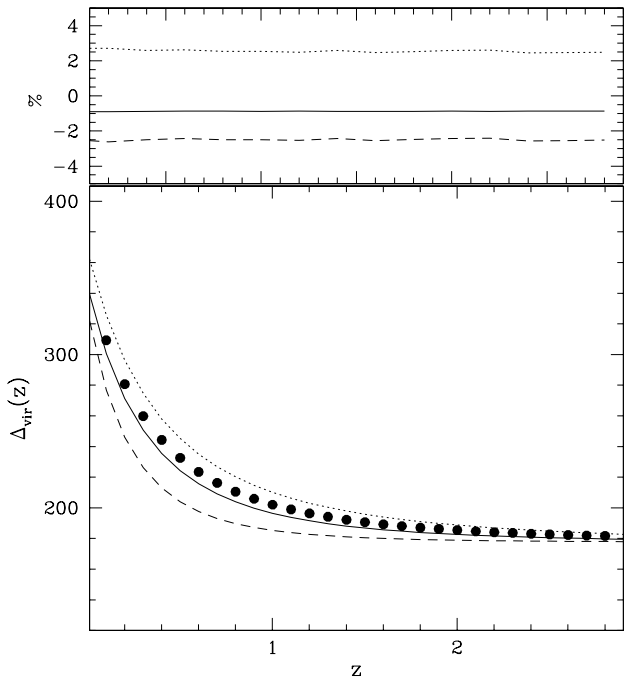}}
  \caption{
{\it Upper Panel:} The deviation $(1-\lambda/\lambda_{\Lambda})\%$
of the collapse factor for various vacuum models with respect to the
$\Lambda$ solution. {\it Bottom Panel:} The density contrast at the
virialization, $\Delta_{vir}$, as a function of redshift. The lines
represent the following cosmological models: (a) $\Lambda_{RGH}$
(solid), (b) $\Lambda_{RGC1}$ (dotted line) and (c) $\Lambda_{RGC2}$
(dashed line). The points represent the traditional $\Lambda$
cosmology.}
        \label{Figcondz3}
 \end{figure}

\begin{table}[h]
\caption[]{Numerical results. The $1^{st}$ column indicates the
vacuum model used. Between column two and four, we present the main
properties of the spherical collapse model
$[\Delta_{vir}(z_{c}),\zeta(z_{t})]$, assuming that galaxy clusters
have collapsed prior to the present time $z_{c}\simeq 0$
($z_{t}\simeq 0.75$). In columns four and
five, we give the same quantities but considering that
clusters have formed (collapsed) prior to epoch of $z_{c}\simeq 1.6$
($z_{t}\simeq 3.2$), in which the most distant cluster has been
found.}
\tabcolsep 5.8pt
\begin{tabular}{ccccc} \hline \hline
Model & $\Delta_{vir}(0)$ &  $\zeta(0.75)$ & $\Delta_{vir}(1.6)$ & $\zeta(3.2)$ \\ \hline
$\Lambda$ & 348.6 & 6.80&190&5.64 \\
$\Lambda_{RGH}$ & 339 & 6.79&184.3&5.62 \\
$\Lambda_{RGC1}$ & 317.6 & 6.66&168.6&5.40 \\
$\Lambda_{RGC2}$ & 368.8 & 6.62&195.2&5.38 \\
\end{tabular}
\end{table}

\subsection{Clustered Vacuum Energy}
In this section we consider a scenario in which the
vacuum energy density on the
scale of galaxy clusters is clustered:
$\rho_{\Lambda s}(R)=\Lambda_{s}(R)/8\pi G$. In such a scenario it could
be possible, on non-linear scales, to have an interaction
between dark matter and dark energy with a different $\gamma$
than the background value.
In the overdensity rest frame
and in the homogeneous case (described before), the dark energy component
flows progressively out of the overdensity \cite{Maor05,david},
and hence energy conservation cannot be applied (especially for large
values of $\gamma$'s) in order to determine the
collapse factor $\lambda$ (along with the virial theorem).
To simplify the inhomogeneous case formalism, we consider the
extreme situation in which the vacuum energy fully clusters along with
the dark matter, avoiding energy non-conservation which was
  examined in ~\cite{Maor05}.

Within this framework, we also assume that the general functional form
that describes the behavior of the vacuum energy density
inside the spherical perturbation obeys a similar equation
as that of eq.(\ref{CRG}):
\begin{equation}
\label{CRGclust}
\Lambda_{s}(R)=\CC_0+
8\pi G \;\frac{\gamma_{s}\,\rho_{ms, t}}{1-\gamma_{s}}\,\left[
\left(\frac{R}{R_{t}}\right)^{-3(1-\gamma_{s})}-1\right]
\end{equation}
or else,
\begin{equation}
\label{CRGclust1}
\Lambda_{s}(y)=\CC_0+
8\pi G \;\frac{\gamma_{s}\,\zeta \rho_{m, t}}{1-\gamma_{s}}\,
\left[y^{-3(1-\gamma_{s})}-1\right] \;,
\end{equation}
where the slope $\gamma_{s}$ is not necessarily equal to the
background $\gamma$.
It is worth noting that
if we leave $\gamma_{s}$ to take also negative values then
there is a critical radius $R_{\star}$, of the spherical overdensity,
in which, for $R<R_{\star}$, the inhomogeneous
vacuum energy density becomes negative $\Lambda_{s}<0$. Thus,
the extra positive pressure ($P_{\Lambda_{s}}>0$) inside the
spherical perturbation implies compression (similar to gravity)
rather than tension (while the opposite is true for $R>R_{\star}$).
With the aid eq.(\ref{CRGclust1}), this critical radius is given by
\begin{equation}
R_{\star}=R_{t}\left[1-\frac{n_{0}(1-\gamma_{s})}
{2\gamma_{s}} \right]^{-\frac{1}{3(1-\gamma_{s})}} \;.
\end{equation}
In the previous equation, we borrowed the definition of
$n_0$ given in (\ref{noo}), where
$\rho_{ms,t}=\rho_{m,t}\zeta=\rho_{m0}\,a_t^{3(1-\gamma)}\zeta$ is
used, in which  $\gamma$ (not $\gamma_s$) is involved  because it
refers to the evolution of the background matter density.

On the other hand, if $\gamma_{s}\ge 0$ then $\Lambda_{s}>0$ (or
$P_{\Lambda_{s}}<0$) for all values of $0<y \le 1$. In this paper we
use 2 different versions of $\Lambda_{s}$, namely
$\gamma_{s}=\gamma=0.002$ (hereafter $\Lambda_{RGC1}$) and
$\gamma_{s}=-0.002$ (hereafter $\Lambda_{RGC2}$). Of course for the
basic cosmological functions [$\Omega_{m}(a)$, $E(a)$ and $D(a)$]
which enter in this vacuum pattern, we utilize the background
$\gamma=0.002$.

Inserting eq.(\ref{CRGclust1}) into eq.(\ref{yddot}), we obtain
\begin{equation}
\label{yddotcl} {\ddot y}=-\frac{H_{t}^{2}\Omega_{m,t}}{2}
\left[ \frac{(1-3\gamma_{s})\zeta}{(1-\gamma_{s})y^{2-3\gamma_{s}}}-
2\left(r-\frac{\gamma_{s} \zeta}{1-\gamma_{s}}\right)y \right]\;.
\end{equation}
In contrast with the homogeneous case, the novelty of the current
approach is that the above differential equation can be solved analytically.
Indeed, due to the fact that the differential eq.(\ref{yddotcl})
is a function only of $y$ we can perform easily the integration
\begin{equation}
{\dot y}^{2}=H_{\rm t}^{2}\Omega_{m, t}\left[P(y,\zeta)+C \right]
\end{equation}
where $C$ is the integration constant and
\begin{equation}
P(y,\zeta)=\frac{\zeta}{(1-\gamma_{s})y^{1-3\gamma_{s}}}+
\left(r-\frac{\gamma_{s} \zeta}{1-\gamma_{s}}\right)y^{2} \;.
\end{equation}
Using now eq.(\ref{xdot1}) we can provide
the basic differential equation for the evolution
of the overdensity perturbations
\begin{equation}
\left(\frac{{\rm d}y}{{\rm d}x}\right)^{2}=
\frac{ P(y,\zeta)+C}
{x^{-1+3\gamma}+rx^{2}I(x)}\;\;,
\end{equation}
where the boundary conditions, $({\rm  d} y/{\rm d} x)=0$
and $y=1$ at $x=1$, imply that $C=-P(1,\zeta)$. Therefore,
the general integral equation which governs the behavior of the
density contrast $\zeta$ at the turn-around epoch, for the RG vacuum
models is
\begin{equation}
\label{zzecl}
\int_{0}^{1} \frac{dy}{\sqrt{P(y,\zeta)+C}}=
\int_{0}^{1} \frac{dx}{\sqrt{x^{-1+3\gamma}+rx^{2}I(x)}} \;.
\end{equation}
Of course, the usual $\CC$CDM cosmology is fully recovered from this model in
the limit $\gamma=\gamma_{s}=0$.

Similarly, as in section 4.1, we again provide a useful fitting formula
for $\zeta$ as well as for $\Delta_{vir}$,
as a function of the cosmological parameters (for more details see
appendix C).



Now with the aid of eq.(\ref{CRGclust}), we can integrate
eq.(\ref{Lener}) in order to derive the potential energy associated
with the vacuum energy inside the spherical overdensity. In
particular, we show here that $U_{\Lambda}$ can be written as a sum
of three components that contribute to the local dynamics
\begin{eqnarray}\label{Lener1}
U_{\Lambda}&=&-\frac{M \Lambda_{0}}{10}R^{2}+4\pi G \frac{\gamma_{s}
M \rho_{ms, t}}{5(1-\gamma_{s})}R^{2}\nonumber\\
&-& 4\pi G \frac{\gamma_{s} M \rho_{ms, t}}
{(1-\gamma_{s})(2+3\gamma_{s})R^{-3(1-\gamma)}_{t}}R^{-1+3\gamma_{s}},
\end{eqnarray}
where $\rho_{ms,t}=\zeta \rho_{m, t}$ (see section 4). In this case,
the algebraic equation which defines the collapse factor is found
from the combination of equations (\ref{energies}), (\ref{Matener})
and (\ref{Lener1}) as follows:
\begin{equation}
\label{collapcl}
q_{2}(\gamma_{s})[n_{0}-f(\gamma_{s})]\lambda^{3}-A(n_{0},\gamma_{s})\lambda
+g(\gamma_{s})\lambda^{3\gamma_{s}}+2q_{1}(\gamma_{s})=0\,,
\end{equation}
where $n_{0}$ is defined in equation (\ref{noo}), with
$$f(\gamma_{s})=\frac{2\gamma_{s}}{1-\gamma_{s}}\;,\;\;\;
g(\gamma_{s})=\frac{10\gamma_{s}q_{2}(\gamma_{s})}{(1-\gamma_{s})(2-3\gamma_{s})}$$
and
$$
A(n_{0},\gamma_{s})=2+n_{0}-f(\gamma_{s})+\frac{g(\gamma_{s})}{q_{2}(\gamma_{s})} \;\;.
$$
Finally, solving
eqs.(\ref{zzecl}) [or using eq.\ref{zetacl1}]
and (\ref{collapcl}), we can estimate
the density contrast at virialization from eq.(\ref{deltavir}).



\section{Comparison among different types of vacuum}
In this section, we investigate in more detail and compare the
spherical collapse model, using different versions of the
$\Lambda(t)$ model (homogeneous or clustered), with that of the
traditional $\Lambda$ cosmology (see appendix A). In figure 2 we
present the density contrast at the turnaround point, $\zeta$, as a
function of the turnaround redshift, $z_t$, for the constant vacuum
$\Lambda$ (solid points), homogeneous vacuum $\Lambda_{RGH}$ (solid
line), clustered vacuum $\Lambda_{RGC1}$ (dotted line) and clustered
vacuum $\Lambda_{RGC2}$ (dashed line). It is obvious that the
$\Lambda$ and $\Lambda_{RGH}$ models are almost indistinguishable,
while $\zeta$ appears to be somewhat lower in the inhomogeneous
(clustered) case. Indeed, we find that $\zeta_{cl}/\zeta_{h} \approx
\pi/3$ at large redshifts.

Solving now eq.(\ref{homvirial}) and eq.(\ref{collapcl}) we
calculate the collapse factor and we find that it lies, in general,
in the range $0.46\le \lambda \le 0.52$, in agreement with previous
studies ( \cite{Bas03}, \cite{Mota04}, \cite{Maor05}, \cite{Wang06},
\cite{Horel05}, \cite{Perc05}, \cite{Basi07}). In the upper panel of
figure 2 we plot the deviation, $(1-\lambda/\lambda_{\Lambda})\%$,
of the collapse factors, $\lambda(z_{c})$, for the current vacuum
models with respect to the $\Lambda$ solution,
$\lambda_{\Lambda}(z_{c})$. It becomes evident that the size of the
cosmic structures which are produced in the $\Lambda_{RGH}$ model
(solid line) is remarkably close to that predicted by the usual
$\Lambda$ cosmology, and therefore the impact of the
vacuum energy on the spherical collapse is very small in the
homogeneous case. This was to be expected.

On the other hand, when considering the effect of the
clustered vacuum energy, the largest positive deviation of the
collapse factor occurs for the $\Lambda_{RGC2}$ model (dotted line),
which implies that this model produces more bound
systems than the concordance $\CC$CDM model. Therefore, within this
vacuum pattern the corresponding cosmic structures should
be located in larger density environments. The opposite
situation holds for the $\Lambda_{RGC1}$ (dashed line) model due to
its negative deviations.

In figure 3 we plot the evolution of the density contrast at
virialization. At very large redshifts, it tends to the Einstein-de
Sitter value ($\Delta_{vir}\sim 18\pi^{2}$), as it should.

In Table 1 we list, for the case of a cluster forming at
$z_{c}\simeq 0$ or at $z_{c}\simeq 1.6$, the following:
(a) the cosmological models and the value of the turn around
redshift; and (b) the virial density $\Delta_{vir}$ at the collapse
time, as well as the density excess of the matter density in the
spherical overdensity, $\zeta$, at the turn around time.

We also verify that the density contrast decreases with the
formation (virialization) redshift $z_{c}$.
The $\Delta_{vir}$ differences among the different vacuum models enter
through $\Omega_{m}(z)$ [see eq. (\ref{aproxf})] as well as
through the assumption about the behavior of the
vacuum inside the spherical overdensity (homogeneous or not).
This feature points that perhaps
the density contrast at virialization can be used as an effective
cosmological tool.

Although, we will investigate in detail such a possibility in a
forthcoming paper, we present here an idea of how to
use observational data to estimate $\Delta_{vir}$. Using existing
catalogs of clusters of galaxies, one should select those clusters
which appear to have a quasi-spherical projected shape\footnote{Not
all apparently spherical clusters are truly spherical since
elongated clusters with their major axis oriented at small angles
along the line of sight will appear spherical in projection, a fact
which is a further source of noise.}, which is the expected shape of
a virialized cosmic structure,
and then derive their virial radius and mass. One can then
easily calculate the observational value of
$\Delta_{vir}$ of a cluster at a redshift $z$, from:
\be
\Delta_{vir} = \frac{3 M_{vir}} {4 \pi \rho_{{\rm crit}, 0} \Omega_{m,0}  (1+z)^3 r_{vir}^3}
\ee
and compare it with the model expectations.
Now, the cluster virial radius can be calculated from the projected
separations of the $N_m$ galaxy members according to (eg., \cite{Merc}):
\begin{equation}
r_{vir} = \frac{\pi}{2}\frac{N_m (N_m-1)}{\sum_{i=1}^{N_m-1} \sum_{j=i+1}^{N_m}
\left[d_L \tan (\delta\theta_{ij})\right]^{-1}}\;\;,
\end{equation}
where $d_L$ is the luminosity distance of the group and
$\delta\theta_{ij}$ is the angular $(i,j)$-pair separation.
Using the observed cluster velocity dispersion, $\sigma_v$, and $r_{vir}$
one can estimate the cluster's virial mass using the virial theorem,
according to:
\begin{equation}
M_{vir}=\frac{3 \sigma_v^2 r_{vir}}{G} + \frac{\Lambda r_{vir}^3}{5 G}\;.
\end{equation}
The second $\Lambda$-based term is negligible,
$\sim 4.7 \times 10^{11} \Omega_{\Lambda} r_{vir}^3 \; M_{\odot}$,
and therefore it does not affect significantly the mass estimates of clusters of galaxies.
Of course, this approach is of a statistical nature,
since there are various observational systematics that enter in the individual
cluster determination of $\Delta_{vir}$, as well as cosmic variance, which however can be
minimized if one averages over a suitable and relatively large sample of clusters
at each redshift interval.

 \begin{figure}[ht]
         \centerline{\includegraphics[width=23pc] {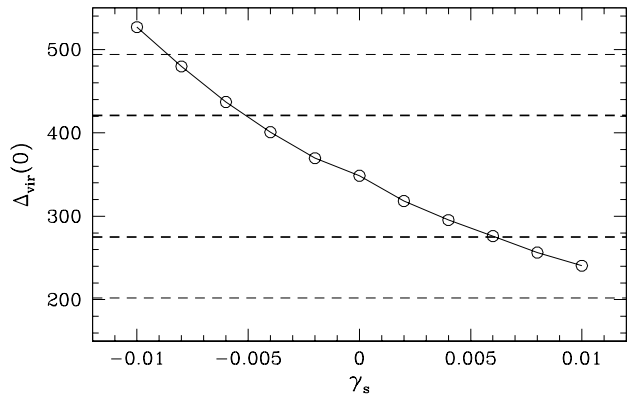}}
  \caption{The present time virial density for the clustered vacuum
  energy model as a function of
$\gamma_{s}$ (open points). The inner and outer dashed lines correspond
to the 1 and 2$\sigma$, observationally determined, virial density limits respectively
($ 202\le \Delta_{vir} \le 494$), based
on a subsample of the 2MASS High Density Contrast group catalog.}
        \label{Figcondz3b}
 \end{figure}

We have applied this methodology to the 2MASS High Density Contrast group catalog
\cite{Crook}, which is
a low-$z$ catalog based on the 2 micron infrared whole-sky survey and
which was constructed by a friends-of-friends algorithm
(eg. \cite{Huchra}) such that the groups
correspond to an overdensity $\delta\rho/\rho \ge 80$.
This catalog was carefully constructed, with respect to other catalogs,
 and it is less prone to projection, interloper contamination and
contamination by the large-scale structures from which galaxies are
accreted to the groups (see \cite{Tov} for a relevant discussion).
We selected only those groups with projected axial ratio $>0.8$ and with at least
16 galaxy members (in order to have a relatively accurate determination of their
shape, velocity dispersion and thus $M_{vir}$) and we are thus left with 7 clusters at
$\langle z \rangle \simeq 0.015$. We clip the lower
and higher $\Delta_{vir}$ outliers, since we do expect systematic effects to be present, and we
derive
a mean value of $\langle \Delta_{vir}\rangle = 348$ and a standard deviation of the distribution
of $\pm 73$ (if we use $N_m>20$ we are left with 6
clusters with $\langle \Delta_{vir}\rangle = 329 \pm 69$). Although, we have derived these
$\Delta_{vir}$ values using the concordance $\Lambda$CDM cosmological model to estimate $d_L$,
there would be no appreciable difference had we used any of the other models, presented in section 2.1
(because of the very small value of $\gamma$ and of the very low redshift of the sample).
Inspecting Table 1 it is evident that the previously
derived observational values are in good agreement with the
theoretical expectations although with the present level of uncertainty we cannot distinguish among
the models.
However, in the case of the clustered vacuum energy model we can put some limits
on the value of $\gamma_{s}$
even with present level of accuracy. As an example, we compare in Figure 4,
the predicted virial density, $\Delta_{vir}$, of the model at the present time
(open points), with the observationally derived 2$\sigma$ range of values,
based on the previously discussed subsample of the 2MASS High Density Contrast group catalog,
and find a consistency for $-0.009 \mincir \gamma_{s} \mincir 0.012$.
In the future we plan to further investigate the model predictions,
using a larger number of clusters spanning also a range of different redshifts,
in an attempt to put stringent constraints on the value of $\gamma_s$.

\section{Conclusions}
In this paper we have studied analytically and numerically the
spherical collapse model in the case of a time varying vacuum,
with $\CC(H)=n_0+n_2\,H^2$, for a spatially flat FLRW geometry.
We find that the amplitude and the shape of
the virial density contrast is affected by the
considered status of the vacuum energy model (homogeneous or
clustered). We verify that in the case where the distribution of
the vacuum energy is clustered the structures produced are
more concentrated (under specific conditions)
with respect to the homogeneous dark energy case.
Finally, 
by comparing the predicted virial density contrast at the present epoch
with a preliminary analysis of a suitable subsample of the 2MASS High
Density Contrast group catalog (at a mean redshift of
$\langle z \rangle \simeq 0.015$), we find that the inhomogeneous vacuum energy models
can be accommodated, at a $2\sigma$ level,
if the vacuum clustering parameter is within the range:
$-0.009\mincir \gamma_{s}\mincir 0.012$.
The latter result points to the direction that perhaps the $\Delta_{vir}$ parameter,
once estimated accurately from observations,
could be used in order to determine the internal physical properties of the vacuum energy.

\vspace {0.5cm}
{\bf Acknowledgments.}
JS has been
supported in part by MEC and FEDER under project FPA2007-66665, by
the Spanish Consolider-Ingenio 2010 program CPAN CSD2007-00042 and
by DIUE/CUR Generalitat de Catalunya under project 2009SGR502. MP
acknowledges funding by Mexican CONACyT grant 2005-49878.

\appendix
\section{The Concordance $\Lambda$ cosmology}
In this appendix we would like to give the reader the opportunity to
appreciate the fact that the $\Lambda_{RG}$ model can be viewed
as an extension of
the concordance $\Lambda$ cosmology.
In particular, the basic cosmological
equations in the $\Lambda_{RG}$ model reduce to those
of the $\Lambda$ cosmology for $\gamma=0$. Below, we present
the main quantities of the $\Lambda$ cosmology:

\begin{itemize}
\item {\it Global Dynamics:} The basic cosmological
equations (see section 2.1) take the following forms:
\begin{equation}
H(t)=\sqrt{\Omega_{\Lambda}}\;H_{0}
\;{\rm coth}\left[\frac{3H_{0}\sqrt{\Omega_{\Lambda}}}{2}\;t\right]
\end{equation}
\begin{equation}
a(t)=\left(\frac{\Omega_{m}}{\Omega_{\Lambda}}\right)^{1/3}
\sinh^{\frac{2}{3}}\left(\frac{3H_{0}\sqrt{\Omega_{\Lambda}} }{2}\; t\right)
\end{equation}
and
\begin{equation}
E^{2}(a)=\frac{H^{2}(a)}{H^{2}_{0}}=\Omega_{\Lambda}+\Omega_{m}a^{-3} \;\;.
\end{equation}

\item {\it The spherical model:}
The basic set of equations here is:
\begin{equation}
{\dot x}^{2}=
H_{t}^{2}\Omega_{m,t} \left[x^{-1}+rx^{2}I(x)\right]
\end{equation}

\begin{equation}
{\ddot y}=-\frac{H_{t}^{2}\Omega_{ m,t}}{2} \left(
\frac{\zeta}{y^{2}}-2r y I(x)\right)
\end{equation}
where $I(x)\equiv 1$ and
\begin{equation}
r=\frac{\Omega_{\Lambda,t}}{\Omega_{m, t}}=r_{0} a^{3}_{t} \;,
\end{equation}
where $r_{0}=\Omega_{\Lambda}/\Omega_{m}$. Therefore, the general
integral equation which governs the behavior of the density contrast
$\zeta$ at the turn around epoch is
\begin{equation}
\int_{0}^{1} \frac{dy}{\sqrt{P(y,\zeta)-P(1,\zeta)}}
    =\frac{{\rm ln}(\sqrt{1+r}+\sqrt{r})^{2/3}}{\sqrt{r}}
\end{equation}
where
\begin{equation}
P(y,\zeta)=\frac{\zeta}{y}+r y^{2} \;.
\end{equation}
Note, that the time needed for a spherical shell to collapse
is twice the turn-around time, $t_{f}\simeq 2t_{t}$.
This implies that:
\begin{equation}
{\rm sinh^{-1}}\left(\sqrt{ r_{0} \;a_{c}^{3}} \right)\simeq 2{\rm
sinh^{-1}}\left(\sqrt{r_{0} \; a_{t}^{3}} \right) \;.
\end{equation}

\item {\it Virial Theorem:}
The virial theorem becomes:
\be
2T+U_{G}-2U_{\Lambda}=0 \;.
\ee
Using now also the energy conservation at the turn around and at the
virial time we derive the following relations:
\be
\frac{1}{2}U_{G,c}+2U_{\Lambda,c}=U_{G,t}+U_{\Lambda,t}
\ee
\begin{equation}
2n_{0} \lambda^{3}-(2+n_{0})\lambda+1=0
\end{equation}
where $\lambda=R_{c}/R_{t}$ is the collapse factor and
\be
n_{0}=\frac{2 \Omega_{\Lambda} a^{3}_{t} }
{\Omega_{m} \zeta} \;.
\ee

\end{itemize}

\section{Roots of a cubic polynomial}
We remind the reader of some basic elements of Algebra pertinent to
our analysis. Given a cubic equation: $\lambda^{3}+a_{1}
\lambda^{2}+a_{2} \lambda+a_{3}=0$, let ${\cal } D$ be the
discriminant:
\begin{equation}
{\cal D}=a_{1}^{2} a_{2}^{2}-4a_{2}^{3}-
4a_{1}^{3}a_{3}-27a_{3}^{2}+18a_{1}a_{2}a_{3}
\end{equation}
and
$$x_{1}=-a_{1}^{3}+\frac{9}{2}a_{1}a_{2}-\frac{27}{2}a_{3} \;,\;\;\;\;\;\;
x_{2}=-\frac{3\sqrt{3 \cal{D}}}{2}\;\;.$$
If ${\cal D}>0$, all roots are real (irreducible case). In that
case $\lambda_{1}$, $\lambda_{2}$ and $\lambda_{3}$ can be written:

\begin{equation}
\lambda_{\mu}=-\frac{a_{1}}{3}-
\frac{2r^{1/3}}{3} {\rm cos} \left[\frac{\theta-(\mu-1)\pi}{3} \right]
\;\;\;\;\;\; \mu=1,2,3
\end{equation}
where $r=\sqrt{x_{1}^{2}+x_{2}^{2}}$ and $\theta={\rm cos^{-1}} (x_{1}/r)$.

Now, we are ready to derive analytically the
exact roots of the basic cubic equation (\ref{homvirial})
having polynomial
parameters: $a_{1}=0$, $a_{2}=-(2+n_{t})/q_{2} n_{c}$ and
$a_{3}=2q_{1}/q_{2}n_{c}$.
Then the discriminant becomes:
\begin{equation}
{\cal D}(n_{t},n_{c})=4\;\frac{(2+n_{t})^{3}-27q^{2}_{1}q_{2}n_{c}}
{q^{3}_{2}n^{3}_{c}} \;\;.
\end{equation}
Of course, in order to obtain physically acceptance solutions we
need to take $n_{t}, n_{c}>0$, which gives ${\cal
D}(n_{t},n_{c})>0$. Therefore, all roots of the cubic equation are
real (irreducible case) but one of them $0\le \lambda_{3}\le 1$
corresponds to expanding shells. It is obvious that for $n_{t},
n_{c} \longrightarrow 0$, the above solution tends to the
Einstein-de Sitter case ($\lambda_{3}\to 0.50$), as it should.

\section{Fitting formulae}
We provide here accurate fitting formulae for
the density contrast at the
turn around redshift and at the epoch of virialization, and which do not
have a simple fully analytical form. These are:
\be \label{zetacl1}
\zeta \simeq \left(\frac{3\pi}{4}\sqrt{1+A_{s}}\right)^{2}
\Omega_{m,t}^{-\omega_{1}+\omega_{2}\Omega_{m,t}-\omega_{3}w(a_{t})}
\ee
where
\be
\label{zeta2}
w(a)=-1-\frac{\gamma a^{3\gamma}} {a^{3\gamma}+\tilde{r}_{0}} \;\;.
\ee
and
\be
\label{aproxf}
\Delta_{vir}(a)\simeq 18\pi^{2}[1+\epsilon \;\Theta^{b}(a)] \;\;.
\ee
where $\Theta(a)=\Omega^{-1}_{m}(a)-1$.

\begin{itemize}
\item {\it Homogeneous Vacuum:}
In this case we have $A_{s}=0$,
$(\omega_{1},\omega_{2},\omega_{3})=(0.79,0.26,0.06)$ and
$$\epsilon=0.40-25\gamma+500\gamma^{2}\;,\;\;b=0.94+50\gamma\;.$$

\item {\it Clustered Vacuum:}
Here we find
\begin{equation}
A_{s}=\left\{ \begin{array}{cc}
       -24.25\gamma_{s}+2125\gamma^{2}_{s}\;\;\;\;
       0\le \gamma_{s}\le 0.01 \\
       29.75\gamma_{s}+2375\gamma^{2}_{s}\;\;\;\;\; -0.01\le \gamma_{s}< 0

       \end{array}
        \right.
\end{equation}

\begin{equation}
(\omega_{1},\omega_{2},\omega_{3})=\left\{ \begin{array}{cc}
       (0.86,0.36,0)\;\;\;\;
       0\le \gamma_{s}\le 0.01 \\
       (0.74,0.16,0)\;\;\;\;\; -0.01\le \gamma_{s}< 0

       \end{array}
        \right.
\end{equation}

\begin{equation}
b=\left\{ \begin{array}{cc}
       0.94+145\gamma_{s}+4.75\times 10^{4}\gamma^{2}_{s} \;\;\;\;
       -0.002\le \gamma_{s}\le 0.01 \\
       0.94+55\gamma_{s} \;\;\;\;\; -0.01\le \gamma_{s}< -0.002

       \end{array}
        \right.
\end{equation}
and
\begin{equation}
\epsilon=\left\{ \begin{array}{cc}
       0.40-65\gamma_{s}-1.25\times 10^{4}\gamma^{2}_{s} \;\;\;\;
       -0.002\le \gamma_{s}\le 0.01 \\
       0.31-86.25\gamma_{s} \;\;\;\;\; -0.01\le \gamma_{s}< -0.002 \;.
       \end{array}
        \right.
\end{equation}

\end {itemize}


\end{document}